# Fractional Order Analysis of the Polytropic Models Applied to Exoplanets

Essam Elkholy[1,2] and Mohamed I. Nouh[2]

[1]Physics Department, College of Science, Northern Border University, Arar, Saudi Arabia
[2]Astronomy Department, National Research Institute of Astronomy and Geophysics, 11421 Helwan, Cairo, Egypt

**Abstract:** Physical conditions deep within planets and exoplanets have yet to be measured directly, but indirect methods can calculate them. The polytropic models are one possible solution to this problem. In the present paper, we assume that the interiors of planets follow a polytropic equation of state. Hydrostatic equilibrium conditions are used to determine the overall structural properties of the constituent matter. In the frame of the conformable fractional derivatives, we use polytropic gas spheres to model the density profiles, pressure profiles, temperature distributions, and the mass-radius relations for the interiors of the initial stage of exoplanets. Planets of single chemical composition were used to study the behavior of the mass-radius relation, pressure distributions, and temperature distribution variation with the fractional parameter. We calculated 72 fractional models for the mass of protoplanets of $1M_J$, $3M_J$, and $10M_J$ ($M_J$ is the mass of Jupiter), and the values of the polytropic index are $n$=0, 0.5, 1, 1.5, and the fractional parameter rang 0.75-1.

Keywords: Analytical Methods- Fractional Derivatives- Polytropes- Exoplanets

1. **Introduction**

Exoplanets, or planets orbiting other stars, are now known to be prevalent in our galaxy. They have a significantly wider physical property than planets in the solar system. Their internal structure can vary from highly puffy gas giants to compact rocky planets with densities as high as iron (Pater, 2015). The explicit temperature dependency in the equation of state is handled differently in the many theories of internal planet matter. So, if the temperature dependence of the equation of state can be neglected with no significant effects, the system can be thought to have a uniform temperature. Models of this sort are commonly referred to as cold planet or zero temperature planet models; as a result, the polytropic equation of state is suitable to model matters inside exoplanets. Until July 18[th], 2022, approximately 5063 exoplanets had been confirmed (https://exoplanetarchive.ipac.caltech.edu/).

Polytropes form the theoretical basis of our understanding of stellar structure and evolution (Chandrasekhar, 1958) and are broadly used in other fields of astrophysics (Horedt, 2004). Polytropic models can calculate known global quantities like mass and radius and conservation equations from the center to the surface. Polytropic equations, for example, have been used to describe the mass and position of planets and moons in solar and satellite systems (Geroyannis and Dallas, 1994), as well as to study globular clusters (e.g., Nguyen and Pedraza, 2013), collapsing molecular clouds and Bok globules (Curry and McKee, 2000), quark stars (Lai and Xu, 2009). Moreover, polytropes have been used to solve stability and oscillation problems and to address relativistic effects in stars, Gleiser and Sowinski (2013), Breysse et al. (2014), Saad et al. (2017), Saad et al. (2021), Geroyannis and Karageorgopoulos (2014).

In recent decades, fractional calculus has had broad applications in physics and engineering, such as quantum physics, wave mechanics, electrical systems, and fractal wave propagation. Various real-life problems are described using fractional differential equations, Stanislavsky (2010) and Herrmann (2014). Fractional derivatives were used by Mathieu et al. (2003) to improve the criterion of thin detection that arises in signal processing. Debnath et al. (2012) generalized the second law of thermodynamics for the Friedmann Universe enclosed by a boundary in the framework of fractional action cosmology.

Many authors implemented fractional-order modeling of astrophysical and space problems. Jamil et al. (2012) developed a dark energy model in fractional action cosmology using a power-law weight function. Analysis of the fractional white dwarf model has been performed by El-Nabulsi (2011), Bayian and Krisch (2015) studied the incompressible gas sphere, and Yousif et al. (2021) examined the fractional isothermal gas sphere using the Taylor series. Nouh and Abdel-Nabulsi (2018) and Abdel-Salam and Nouh (2020) investigated the polytropic gas sphere using power series expansion.

The present paper uses fractional polytropes to simulate physical variables such as radius, mass, density, temperature, and pressure inside the exoplanets. We attempt to calculate precise values of the zero of the Emden function for a given polytropic index and fractional parameter to calculate the radius and mass accurately. We will use single fractional polytropes to investigate the interior structure of exoplanets. We assume isolated spherical gaseous protoplanets with the solar composition of gas generated by gravitational instability and masses ranging from 0.3 to 10 $M_J$ (Paul et al., 2021); it is worth noting that this mass range encompasses most of the discovered exoplanets (Helled and Schubert, 2008). During their early stages, protoplanets contract quasi-statically, the ideal gas law holds good, and the only energy source is the gravitational contraction (Paul et al., 2020; Paul et al., 2021). We suppose that such a protoplanet is in a stable state in which the polytropic equation of state holds well (Paul et al., 2014). Using symbolic manipulation in MATHEMATICA 12, accelerated analytical expressions for the physical quantities will be generated.

The structure of the paper is as follows: Section 2 is devoted to the principles of the conformable derivative, section 3 deals with the fractional polytropic equation and the physical parameters of the polytrope, section 4 gives the computational method used for calculations, in section 5 we present the results of the analysis, and in section 6 we outlined the concluding remarks.

## 2. Conformable Fractional Derivatives (CFD)

The conformable fractional derivative (CFD) uses the limits in the form (Khalil et al., 2014)

$$D^{\alpha} f(t) = \lim_{\varepsilon \to 0} \frac{f(t + \varepsilon t^{1-\alpha}) - f(t)}{\varepsilon} \quad \forall t > 0, \alpha \in (0,1] \tag{1}$$

$$f^{(\alpha)}(0) = \lim_{t \to 0^+} f^{(\alpha)}(t). \tag{2}$$

Here $f^{(\alpha)}(0)$ is not defined. When $\alpha = 1$ this fractional derivative reduces to the ordinary derivative. The conformable fractional derivative has the following properties:

$$D^{\alpha} t^p = p t^{p-\alpha}, \ p \in \mathbb{R}, \ D^{\alpha} c = 0, \quad \forall f(t) = c \tag{3}$$

$$D^{\alpha}(af + bg) = aD^{\alpha}f + bD^{\alpha}g, \ \forall \, a, b \in \mathbb{R}, \tag{4}$$

$$D^{\alpha}(fg) = fD^{\alpha}g + fD^{\alpha}g \tag{5}$$

$$D^{\alpha} f(g) = \frac{df}{dg} D^{\alpha} g, \tag{6}$$

$$D^{\alpha} f(t) = t^{1-\alpha} \frac{df}{dg} \tag{7}$$

where $f$, $g$ are two $\alpha$-differentiable functions and $c$ is an arbitrary constant. Equations (5) to (6) are proved by Khalil et al. (2014). The conformable fractional derivative of some functions

$$D^\alpha e^{ct} = ct^{1-\alpha}e^{ct}, \quad D^\alpha \sin(ct) = ct^{1-\alpha}\cos(ct), \quad D^\alpha \cos(ct) = -ct^{1-\alpha}\sin(ct), \qquad (8)$$

$$D^\alpha e^{ct^\alpha} = c\alpha e^{ct^\alpha}, \quad D^\alpha \sin(ct^\alpha) = c\alpha \cos(ct^\alpha), \quad D^\alpha \cos(ct^\alpha) = -c\alpha \sin(ct^\alpha). \qquad (9)$$

### 3. The Fractional Polytropic Model

Polytropes use the equation of state $P = K\rho^{1+1/n}$ with a constant $K$ (proportional to the gas entropy) and the polytropic index $n$ to derive hydrostatic pressure $P$ and mass density. Any gravitating body where this simple EOS can be applied can be considered a polytrope.

The fractional form of equations of mass conservation and hydrostatic equilibrium is given by (Abdel-Salam and Nouh, 2020)

$$\frac{d^\alpha M(r)}{dr^\alpha} = 4\pi r^{2\alpha}\rho, \qquad (10)$$

and

$$\frac{d^\alpha P(r)}{dr^\alpha} = -\frac{GM(r)}{r^{2\alpha}}\rho. \qquad (11)$$

Rearrange Equation (11) and perform the first fractional derivative we get

$$\frac{d^\alpha}{dr^\alpha}\left(\frac{r^{2\alpha}}{\rho}\frac{d^\alpha P(r)}{dr^\alpha}\right) = -G\frac{d^\alpha M(r)}{dr^\alpha}. \qquad (12)$$

Inserting Equation (10) into Equation (12), we obtain

$$\frac{1}{r^{2\alpha}}\frac{d^\alpha}{dr^\alpha}\left(\frac{r^{2\alpha}}{\rho}\frac{d^\alpha P(r)}{dr^\alpha}\right) = -4\pi G\rho. \qquad (13)$$

If $\rho$ and $\rho_c$ denote the density and the central density, the Emden function ($u$) could be defined as

$$u = \left(\frac{\rho}{\rho_c}\right)^{1/n} \qquad (14)$$

Define the dimensionless variable $x$ as

$$x^\alpha = \frac{r^\alpha}{a}. \tag{15}$$

Inserting Equations (10) and (14) in Equation (13), we get

$$\frac{K}{(ax^\alpha)^2} \frac{d^\alpha}{d(ax^\alpha)} \left[ \frac{(ax^\alpha)^2}{\rho_c u^n} \frac{d^\alpha (\rho_c u^n)^{1+\frac{1}{n}}}{d(ax^\alpha)} \right] = -4\pi G \rho_c u^n. \tag{16}$$

Take the fractional derivative of $(u)$

$$\frac{d^\alpha}{dx^\alpha} u^{n+1} = (n+1) u^n \frac{d^\alpha u}{dx^\alpha}. \tag{17}$$

Inserting Equation (17) in Equation (16) and rearranging, we get

$$\frac{K(n+1)\rho_c^{\frac{1}{n}-1}}{4\pi G a^2} \frac{1}{x^{2\alpha}} \frac{d^\alpha}{d x^\alpha} \left( x^{2\alpha} \frac{d^\alpha u}{d x^\alpha} \right) = -u^n. \tag{18}$$

Now by taking

$$a^2 = \frac{K(n+1)\rho_c^{\frac{1}{n}-1}}{4\pi G}, \tag{19}$$

then the Lane-Emden equation in its fractional form is given by

$$\frac{1}{x^{2\alpha}} \frac{d^\alpha}{d x^\alpha} \left( x^{2\alpha} \frac{d^\alpha u}{d x^\alpha} \right) = -u^n. \tag{20}$$

### 4. Computational Method

Write the Equation (20) in the form

$$x^{-2\alpha} D_x^\alpha \left( x^{2\alpha} D_x^\alpha \right) u + u^n = 0, \tag{21}$$

by putting $X = x^\alpha$, the Emden function could be computed from the series form (Abdel-Salam and Nouh, 2020)

$$u(X) = \sum_{m=0}^{\infty} A_m X^m, \tag{22}$$

where the series coefficients will be calculated using the recurrence relations

$$A_{k+2} = -\frac{Q_k}{\alpha^2 (k+2)(k+3)}, \quad \forall \, k \geq 2 \tag{23}$$

and

$$Q_m = \frac{1}{m! A_0} \sum_{i=1}^{m} (m-1)!(in-m+i) A_i Q_{m-i}, \quad \forall \, m \geq 1, \tag{24}$$

Equation (21) has exact solutions only for the polytopes with n=0, 1, and 5 given by

$$\begin{aligned} y(x) &= 1 - \frac{1}{6}\left(\frac{x^\alpha}{\alpha}\right)^2, \\ y(x) &= \left(\frac{x^\alpha}{\alpha}\right)^{-1} \sin\left(\frac{x^\alpha}{\alpha}\right), \\ y(x) &= \left(1 + \frac{1}{3}\left(\frac{x^\alpha}{\alpha}\right)^2\right)^{-\frac{1}{2}}. \end{aligned} \tag{25}$$

The mass contained in a radius $r$ is provided by

$$M(r^\alpha) = \int_0^r 4\pi r^{2\alpha} \rho \, dr^\alpha . \tag{26}$$

Inserting Equations (15) and (14) for $\rho$ and $r^\alpha$ we found

$$M(x^\alpha) = 4\pi a^3 \rho_c \int_0^x x^{2\alpha} u^n \, dx^\alpha \tag{27}$$

by substituting Equation (20) for the Emden function $u^n$, we get

$$M(x^\alpha) = 4\pi \left[\frac{K(n+1)}{4\pi G}\right]^{\frac{3}{2}} \rho_c^{\frac{3-n}{2n}} \left[-\left(x^{2\alpha} \frac{d^\alpha u}{d x^\alpha}\right)\right]_{x=x_1} . \tag{28}$$

The radius of the polytropic gas sphere is given by

$$R^{\alpha} = a\, x_1^{\alpha},$$

where $x_1^{\alpha}$ is the first zero of the Lane-Emden function. Then the radius is given by

$$R^{\alpha} = \left[\frac{K(n+1)}{4\pi G}\right]^{\frac{1}{2}} \rho_c^{\frac{1-n}{2n}} x_1^{\alpha}. \tag{29}$$

The pressure and the temperature distributions could be calculated from

$$\begin{aligned} P &= P_c\, u^{n+1}, \\ T &= T_c\, u^n \end{aligned} \tag{30}$$

### 5. Results and Discussion

To determine the structure of the conformable polytropic gas sphere, we used the analytical solution of Equation (21) with the two recurrence relations, Equations (23-24). We elaborated a MATHEMATICA code to calculate the series coefficients, radii, densities, pressures, temperatures, and masses of the polytropic gas spheres for the range of the polytropic index $0 \leq n < 3$. To explore the properties of the models, we computed 72 polytropic models for *n* and $\alpha$ values listed in Table 1.

It is worth noting that the Emden function computed using the power series without applying acceleration techniques is limited to the interior points ($0 \leq x \leq 1$). To reach the surface of the polytrope, we implemented the accelerating scheme proposed by Nouh (2004) to accelerate the series convergence. As an example, we depict in Figure 1 the Emden function calculated for n=2 and $\alpha = 0.95$; the dashed line is for the power series solution without acceleration, and the solid line is for the calculation with the accelerated power series.

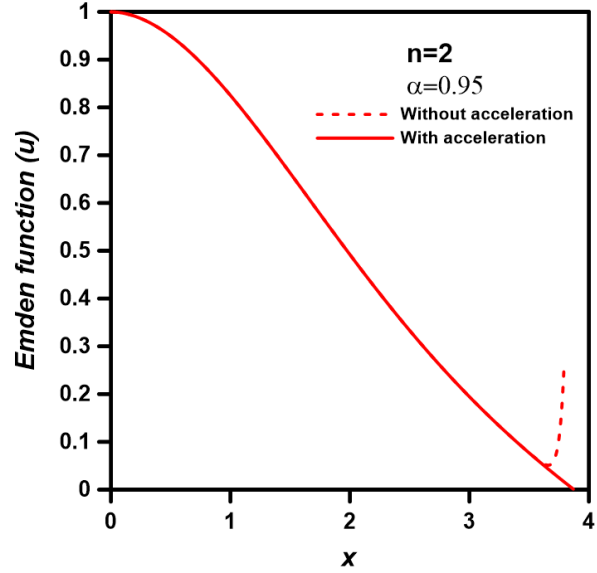

Figure 1: The distribution of Emden function for conformable polytrope with n=2 and $\alpha = 0.95$. The dashed line represents the unaccelerated solution and the solid line for the accelerated solution.

For a given polytropic index *n* and a fractional parameter $\alpha$, the first step of the calculations is to compute the zeroth of the Emden functions (analog to the radius of the gas sphere). In Table 1, we listed the calculated zeroth for the range of the *n*=0 (0.5) 3 and $\alpha$ =0.75 (0.10) 1; it is well noticed that the radius of the gas sphere is decreased as the fractional parameter $\alpha$ decreased. This decrease means that the volume of the fractional gas sphere would be smaller than the integer one. From the table, we can calculate the percentage $x_1(\alpha = 0.75) / x_1(\alpha = 1)$ of the radii of the fractional gas spheres with $\alpha$ =0.75 to that of the integer one ($\alpha$ =1) have the following values: for n=0 the radius of the fractional sphere is reduced to 76% of its integer value, while for n=0.5 is 75%, for n=1 is 73%, for n=1.5 is 68%, for n=2 is 61%, for n=2.5 is 53% , and finally for n=3 is 43%. These results indicate that the volume of the fractional gas sphere is smaller than that of the integer one, and this decrease in the volume is directly proportional to the polytropic index.

Table 1: The zero ($x_1$) of the Emden function for different fractional polytropic models.

| n | $x_1$ | | | | | |
|---|---|---|---|---|---|---|
| | $\alpha=1$ | $\alpha=0.95$ | $\alpha=0.9$ | $\alpha=0.85$ | $\alpha=0.8$ | $\alpha=0.75$ |
| 0 | 2.44 | 2.36 | 2.27 | 2.15 | 2.01 | 1.85 |
| 0.5 | 2.75 | 2.65 | 2.52 | 2.40 | 2.24 | 2.06 |
| 1 | 3.14 | 3.0 | 2.85 | 2.68 | 2.49 | 2.28 |
| 1.5 | 3.65 | 3.41 | 3.19 | 2.96 | 2.73 | 2.48 |
| 2 | 4.35 | 3.87 | 3.54 | 3.24 | 2.96 | 2.67 |
| 2.5 | 5.35 | 4.37 | 3.88 | 3.50 | 3.16 | 2.83 |
| 3 | 6.89 | 4.87 | 4.2 | 3.74 | 3.35 | 2.98 |

In the present study, we consider isolated spherical gaseous protoplanets with the solar composition of the gas with masses ranging from 0.3 to 10 $M_J$. It is worth mentioning that the mass range includes most known exoplanets in their early phases (Helled and Schubert, 2008). We assume that the protoplanet is stable with a well-fitting polytropic equation of state. We modeled the fractional polytropes having solar chemical compositions with the parameters listed in Table 2 for the range of the fractional parameter $\alpha=0.75-1$. The central values of the densities, the pressures, and the temperatures are taken from Paul et al. (2014), based on the study of Helled and Schubert (2008).

Table 2: Central values of the density, pressure, and temperature of protoplanetary masses and radii of the polytrope with index *n* (Helled and Schubert, 2008).

| $M/M_J$ | $R \times 10^{12}$ (cm) | $\rho_c \times 10^{-9}$ (gm cm$^{-3}$) | | | | $P_c$ (dyne cm$^{-2}$) | | | | $T_c$ (K) | | | |
|---|---|---|---|---|---|---|---|---|---|---|---|---|---|
| | | n=0 | n=0.5 | n=1 | n=1.5 | n=0 | n=0.5 | n=1 | n=1.5 | n=0 | n=0.5 | n=1 | n=1.5 |
| 1 | 5.3 | 3.04 | 5.59 | 10.02 | 18.23 | 36.31 | 709.10 | 1181.87 | 1632.27 | 318.83 | 278.09 | 318.83 | 343.37 |
| 3 | 7.8 | 2.87 | 5.26 | 9.42 | 17.16 | 69.83 | 1360.43 | 2276.51 | 4031.02 | 649.92 | 556.87 | 649.92 | 699.96 |
| 10 | 11.0 | 3.41 | 6.25 | 11.20 | 22.40 | 196.17 | 3821.57 | 12738.48 | 8796.85 | 1536.17 | 1339.86 | 1536.18 | 1654.44 |

The distributions of the physical parameters of the fractional polytrope like mass, radius, pressure, temperature, and density could be computed from Equations (28-30). Using the zeros of the Emden function ($x_1$) listed in Table 1 for each (n, $\alpha$) pair, we calculated 72 conformable polytropic models. In Figures (2-4), we plotted the pressure profile, the mass-radius relation, and the temperature distribution for an exoplanet with a mass of $3M_J$ and the polytropic indices n=0, 0.5, 1, and 1.5. The values of central densities, central pressures, and central temperatures are taken from Table 2. For the polytropic indices $0 \leq n \leq 1.5$, the fractional gas sphere has a smaller pressure than the integer one. In Figure 3, we plot the mass-radius relations; the behavior is opposite to the pressure profile; the sphere's mass increases with the fractional parameter. The temperature will behave similarly to the pressure (as shown in Figure 4) since temperature and pressure are related to the central values through the Emden function.

The pressure profiles for protoplanets with masses $1M_J$, $3M_J$, and $10M_J$ are plotted in Figure 5; the fractional models are computed for the polytropic index n=0.5 (upper panel) and n=1.5 (lower panel). It is shown that from Figure 5, the pressure for both integer and fractional models of the protoplanets increases with increasing masses. The pressure near the protoplanets' center is nearly identical for the integer and fractional models. Another important notice is that the fractional models have more minor effects for the polytrope with n=1.5 than for the polytrope with n=0.5. Figure 6 shows the temperature distributions inside protoplanets with $1M_J$, $3M_J$, and $10M_J$ and polytopic indices n=0.5 and 1.5. As the mass of the protoplanet increase, the temperature increases. Our findings for the integer models ($\alpha$ =1) are in reasonable agreement with the calculations of Paul and Bhattacharjee (2013) and Paul et al. (2014).

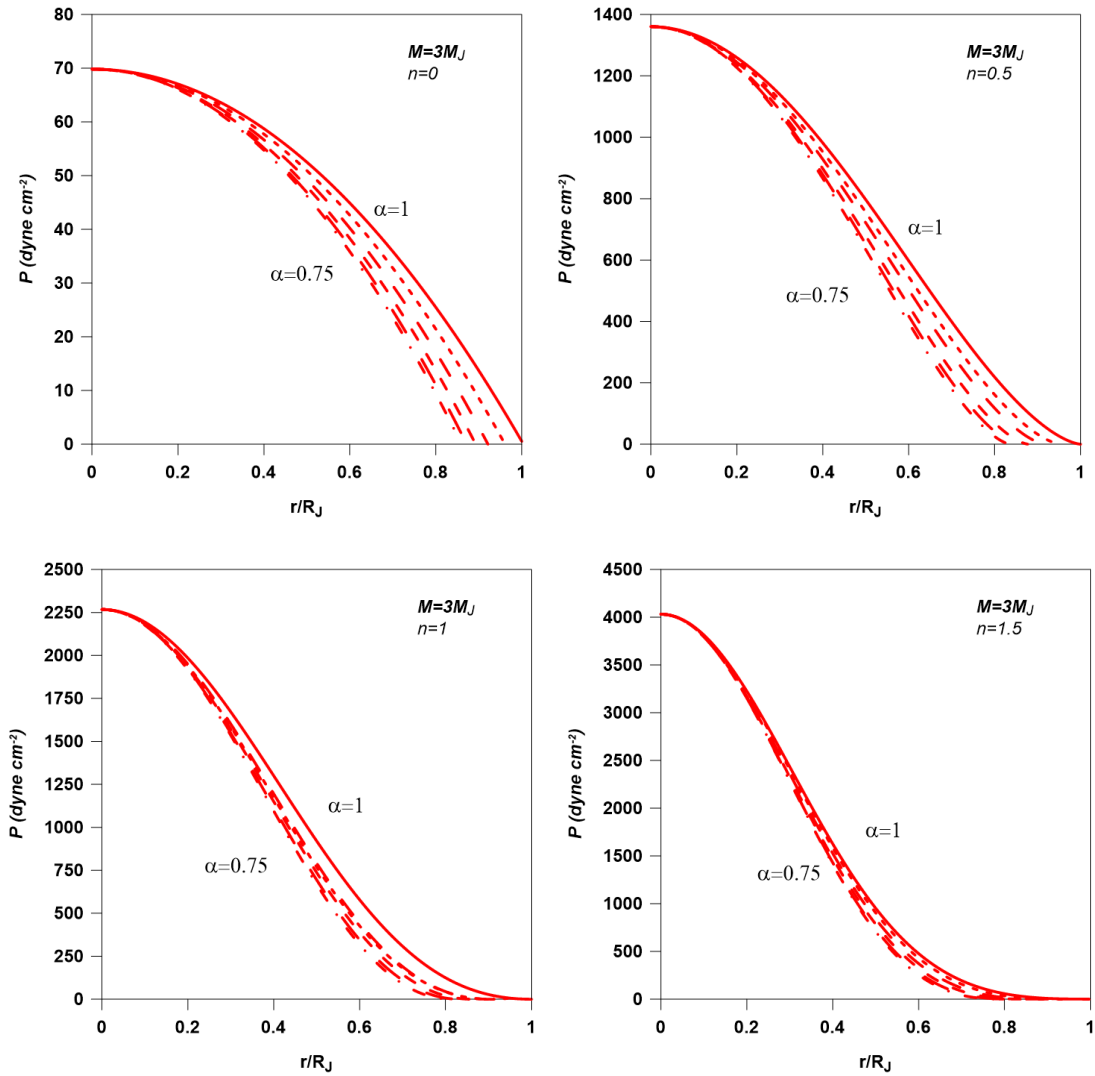

Figure 2: The pressure profile of fractional polytropic gas spheres. The polytropic index ranged from n=0-1.5, and the fractional parameter ranged from 0.75-1.

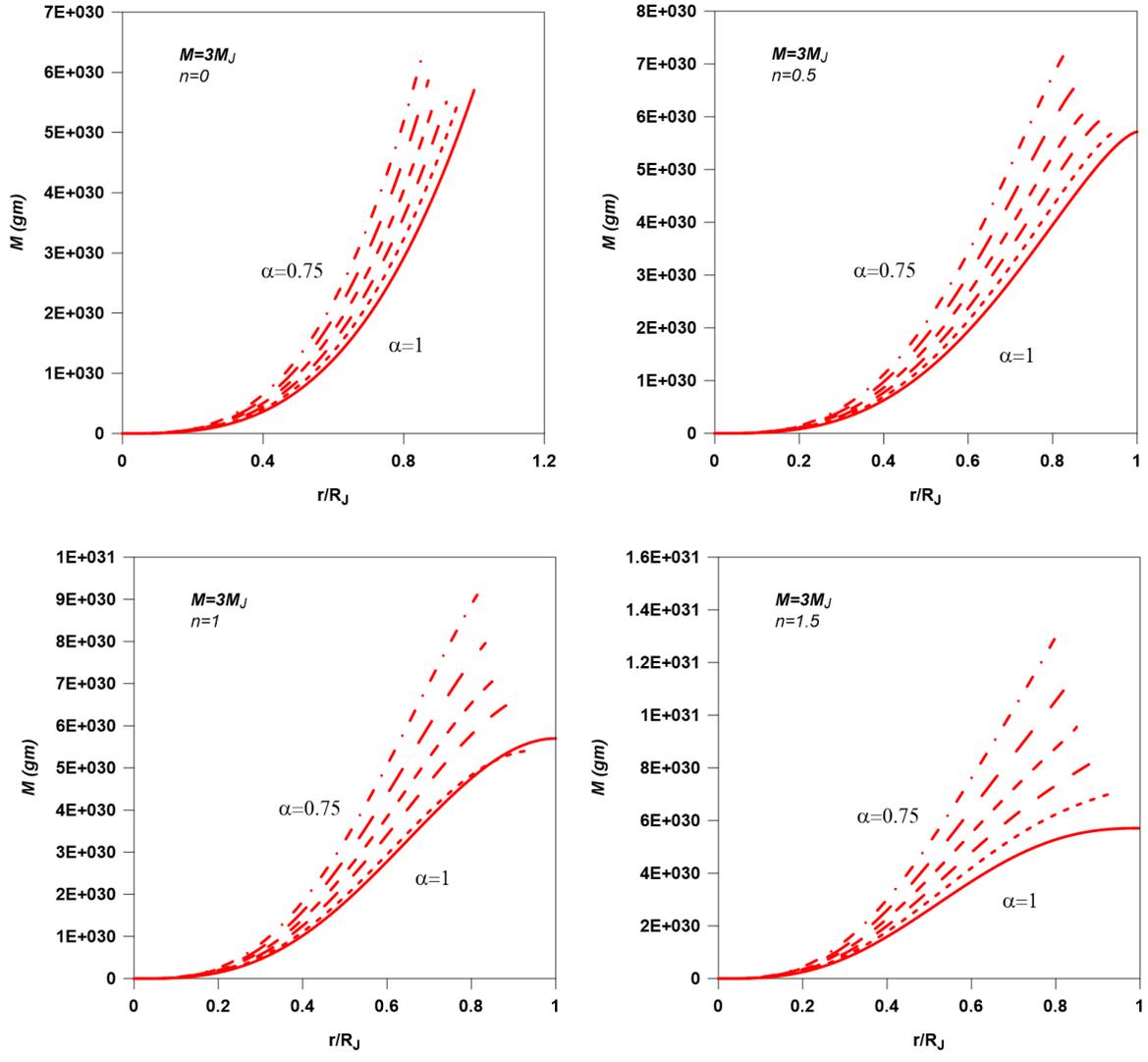

Figure 3: The mass-radius relations for fractional polytropic gas spheres. The polytropic index ranged from n=0-1.5, and the fractional parameter ranged from 0.75-1.

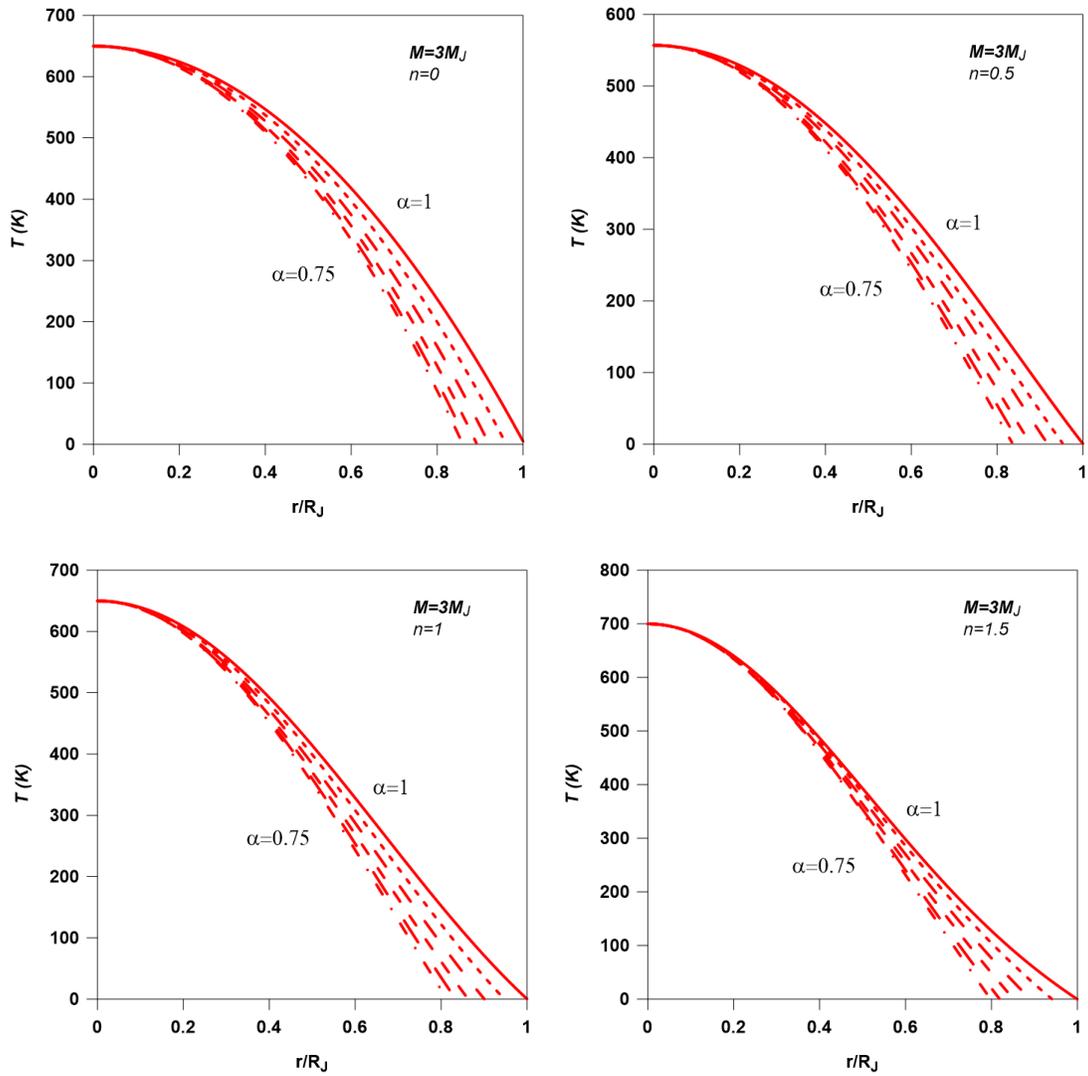

Figure 4: The temperature distribution for fractional polytropic gas spheres. The polytropic index ranged from n=0-1.5, and the fractional parameter ranged from 0.75-1.

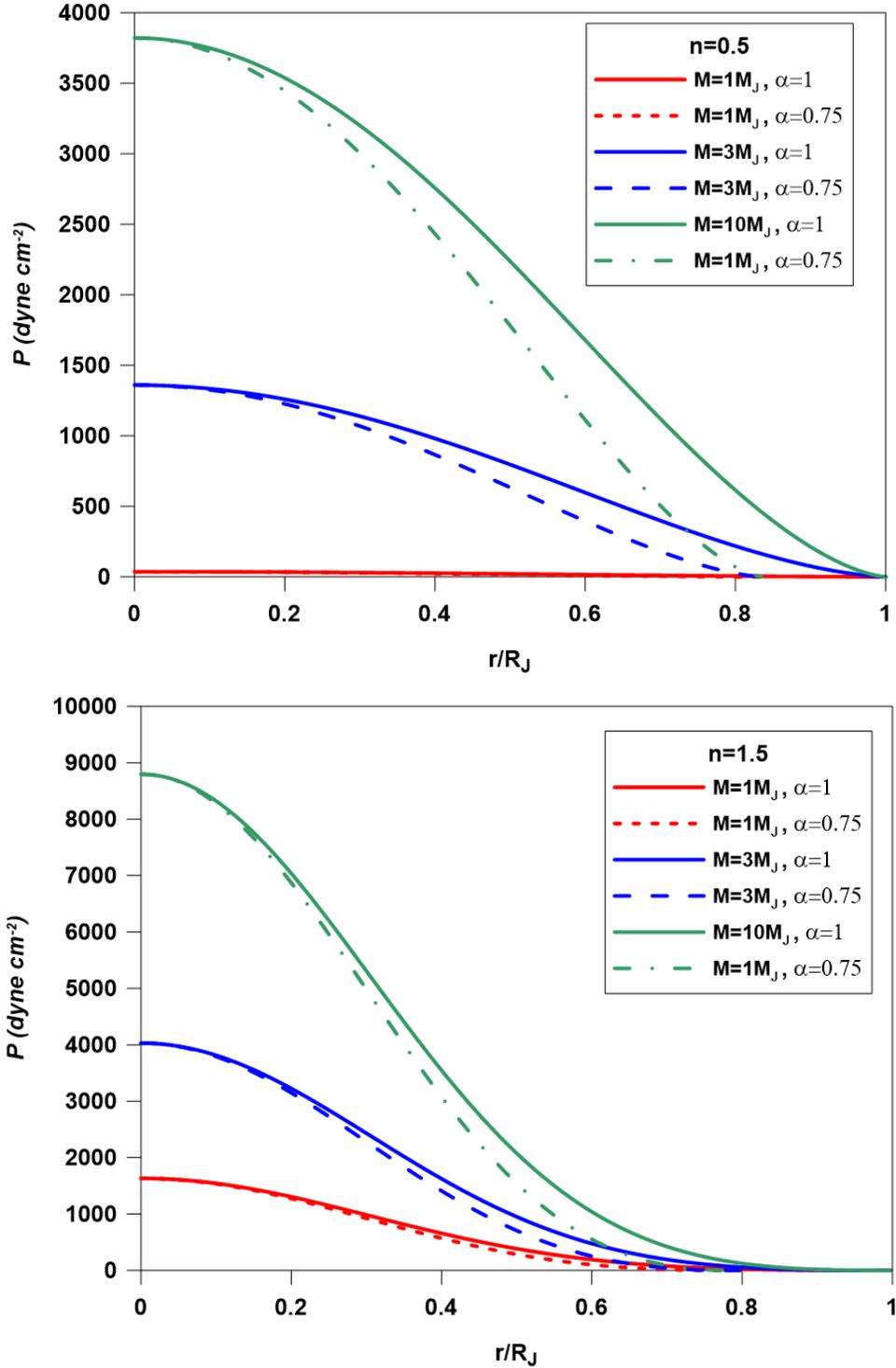

Figure 5: The pressure profile of the fractional polytrope with n=0.5 (upper panel) and n=1.5 (lower panel), computed for the masses of 1$M_J$ (red lines), 3$M_J$ (blue lines), and 10$M_J$ (green lines). The solid lines are for the fractional parameter $\alpha = 1$, and the dashed lines are for $\alpha = 0.75$.

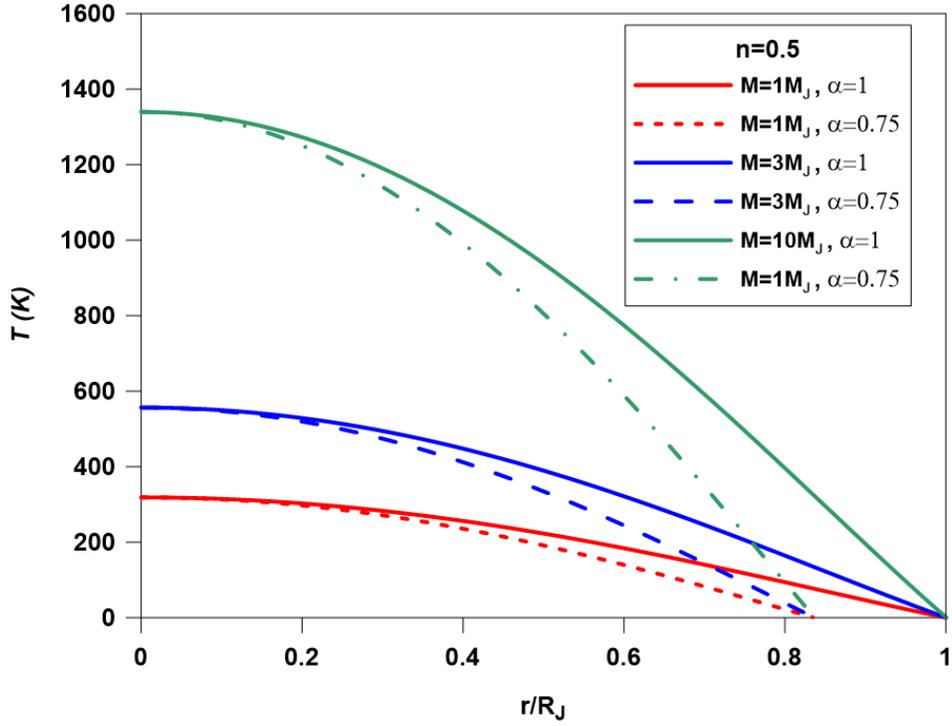

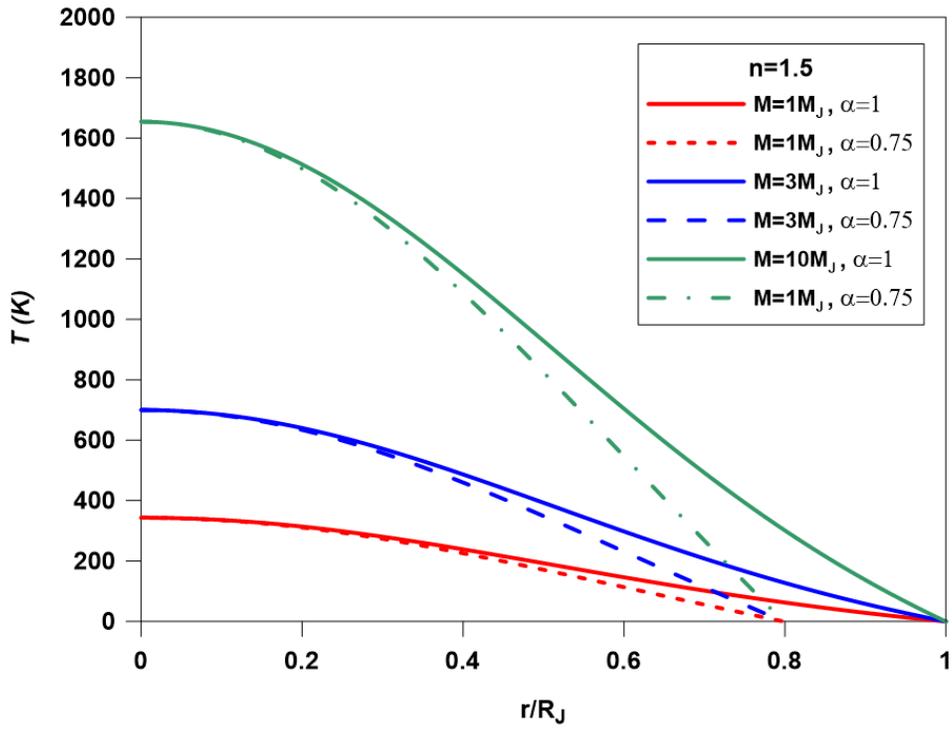

Figure 6: The temperature distribution for fractional polytrope with n=0.5 (upper panel) and n=1.5 (lower panel), computed for the masses of 1$M_J$ (red lines), 3$M_J$ (blue lines), and 10$M_J$ (green lines). The solid lines are for the fractional parameter $\alpha = 1$, and the dashed lines are for $\alpha = 0.75$.

# 5  Conclusion

In the present paper, we assumed isolated spherical gaseous protoplanets with the solar composition of the gas with masses ranging from 0.3 to 10 $M_J$. The protoplanet is stable with a well-fitting polytropic equation of state. We computed conformable fractional polytropic models for the initial stage of protoplanets. We implemented the accelerated power series expansion to solve the conformable fraction LE equation. The fractional models are computed for masses $1M_J$, $3M_J$, and $10M_J$, polytropic indices n=0, 0.5, 1, and 1.5, and the fractional parameters range 0.75-1. The results could be summarized in the following points:

- For the fractional models with polytropic indices $0 \leq n \leq 1.5$, the gas sphere has a smaller pressure than the integer one, and the temperature behaves as the pressure.
- The behavior of the mass-radius relation is opposite to the pressure profile; the sphere's mass increases with the fractional parameter.
- The fractional gas sphere's radius (i.e., volume) is smaller than that of the integer one. This decrease in the volume of the gas sphere is directly proportional to the polytropic index.
- For each mass track, the pressure profiles for protoplanets for both integer and fractional models show an increase with increasing masses. The pressure near the protoplanets' center is nearly identical for the integer and fractional models. Another important notice is that the fractional models have more negligible effects for the polytrope with n=1.5 than for the polytrope with n=0.5. The temperature of the protoplanet rises as the protoplanet's mass rises.
- Our results for integer models ($\alpha$ =1) are in reasonable agreement with earlier investigations of Paul and Bhattacharjee (2013) and Paul et al. (2014).